\documentclass[letterpaper]{article} 
\usepackage{aaai2026}  
\usepackage{times}  
\usepackage{helvet}  
\usepackage{courier}  
\usepackage[hyphens]{url}  
\usepackage{graphicx} 
\usepackage{natbib}  
\usepackage{caption} 
\usepackage{algorithm}
\usepackage{algorithmic}
\usepackage{graphicx}
\usepackage{subcaption}
\usepackage{float}
\usepackage{booktabs}
\usepackage{enumitem}
\setlist[enumerate,1]{label=\arabic*), leftmargin=1.5em}
\usepackage{amsthm}
\usepackage{amsmath}

\usepackage{tikz}
\usetikzlibrary{positioning}

\usepackage[none]{hyphenat}
\usetikzlibrary{arrows.meta,positioning,shapes.geometric} 

\usepackage{tikz}
\usetikzlibrary{decorations.pathreplacing,calc,positioning}
\usepackage{newfloat}
\usepackage{listings}
\DeclareCaptionStyle{ruled}{labelfont=normalfont,labelsep=colon,strut=off} 
\floatstyle{ruled}
\newfloat{listing}{tb}{lst}{}
\floatname{listing}{Listing}
\title{$\Delta_1$–LLM: Symbolic–Neural Integration for Credible and Explainable Reasoning}

\author{
    Yang Xu\textsuperscript{\rm 1,3},
    Jun Liu\textsuperscript{\rm 2,3}\thanks{*~Corresponding author},
    Shuwei Chen\textsuperscript{\rm 1,3},
    Chris Nugent\textsuperscript{\rm 2},
    Hailing Guo\textsuperscript{\rm 1,3}
}
\affiliations{
    \textsuperscript{\rm 1}School of Mathematics, Southwest Jiaotong University, Chengdu 611756, China\\
    \textsuperscript{\rm 2}School of Computing, Ulster University, Belfast, Northern Ireland, UK\\
    \textsuperscript{\rm 3}National–Local Joint Engineering Laboratory of System Credibility Automatic Verification, Southwest Jiaotong University, Chengdu 611756, Sichuan, China\\
    xuyang@swjtu.edu.cn, j.liu@ulster.ac.uk, swchen@swjtu.edu.cn, cd.nugent@ulster.ac.uk,
    guo\_hl@my.swjtu.edu.cn
}
\begin{document}

\maketitle

\begin{abstract}
Neuro-symbolic reasoning increasingly demands frameworks that unite the formal rigor of logic with the interpretability of large language models (LLMs). We introduce an end-to-end explainability-by-construction pipeline integrating the Automated Theorem Generator $\Delta_{1}$ based on the full triangular standard contradiction (FTSC) with LLMs. $\Delta_{1}$ deterministically constructs minimal unsatisfiable clause sets and complete theorems in polynomial time, ensuring both soundness and minimality by construction. The LLM layer verbalizes each theorem and proof trace into coherent natural-language explanations and actionable insights. Empirical studies across health care, compliance, and regulatory domains show that $\Delta_{1}$ + LLM enables interpretable, auditable, and domain-aligned reasoning. This work advances the convergence of logic, language, and learning, positioning constructive theorem generation as a principled foundation for neuro-symbolic explainable AI.
\end{abstract}


\section{Introduction}

Recent advances in neuro-symbolic reasoning reveal a growing need for systems that unify the formal rigor of symbolic logic with the interpretability of large language models (LLMs). Symbolic approaches provide provable soundness and transparent inference structure but lack linguistic expressivity; neural models offer fluency and adaptability yet no guarantees of correctness. Bridging these paradigms calls for architectures where logical validity and natural-language explainability coexist by design.

Building on the most recent work of deterministic Automated Theorem Generator \textbf{$\Delta_{1}$}~\cite{Xu2025atg}, grounded in the formal structure of \emph{Full Triangular Standard Contradiction} (FTSC)~\cite{Xu2026contradiction}, this paper introduces an integrated framework that combines $\Delta_{1}$ with LLMs for interpretable reasoning. $\Delta_{1}$ generates sound and complete theorems without search or stochastic guidance, while the LLM verbalizes proofs, ranks explanatory salience, and produces coherent rationales. Together they establish an \emph{explainability-by-construction} pipeline: logical precision from $\Delta_{1}$, semantic intelligibility from the LLM.

Through case studies in health care, compliance auditing, regulatory conformance, and contract analysis, we show that $\Delta_{1}$ + LLM identifies minimal contradictions and transforms them into auditable, domain-aligned explanations. This integration advances the goal of trustworthy, human-centered neuro-symbolic AI, reasoning that is both verifiable by machines and comprehensible to humans.

\paragraph{Contributions:} This work contributes:
\begin{itemize}
    \item A neuro-symbolic integration pipeline that couples the deterministic theorem generation of $\Delta_{1}$ with the interpretive capabilities of LLMs.
    \item Theoretical justification and empirical validation showing that $\Delta_{1}$~+~LLM delivers both formal soundness and human-level interpretability across high-stakes reasoning domains.
    \item A demonstration of \emph{explainability by construction}, where logical derivations inherently produce transparent, auditable natural-language explanations.
\end{itemize}

\section{Related Works}
\label{sec:related}
\paragraph{Logic, Learning, and Neuro-Symbolic Reasoning.}
Automated reasoning has progressed from early symbolic systems based on explicit rule manipulation to contemporary neuro-symbolic approaches that unify learning and inference. The proposed $\Delta_{1}$ + LLM framework builds on this lineage, connecting to work in theorem proving, minimal unsatisfiability, and closure generation.

The integration of logic and learning traces back to symbolic AI~\cite{mccarthy1959programs}, inductive logic programming (ILP)~\cite{MUGGLETON1994629}, and modern neuro-symbolic systems~\cite{manhaeve2021,YU2023105}. Recent models such as Logic-LM++~\cite{kirtania2024}, VERUS-LM~\cite{callewaert2025}, and MuSLR~\cite{xu2026MUSLR} embed symbolic constraints into transformers to enhance consistency, though they lack formal soundness. Hybrid reasoning methods like RLLF~\cite{nguyen2023} and Logic-of-Thought~\cite{liu2025LoT}, along with Logic-LM~\cite{pan2023logiclm} and DeepProbLog~\cite{manhaeve2018deepproblog}, combine neural and logical inference through structured prompting or probabilistic links. While LLMs show potential in proof synthesis~\cite{polu2022formal,wu2023autoformalization}, their reasoning remains stochastic and unverifiable. As surveys note~\cite{colelough2025,fang2024}, robust neuro-symbolic systems require sound symbolic cores.

The $\Delta_{1}$ + LLM framework meets this need by coupling a \emph{constructive theorem generator} with a \emph{language-based evaluator}, achieving correctness-by-construction and human-centered interpretability—marking, to our knowledge, the first formally guaranteed, linguistically interpretable discovery engine bridging symbolic and natural-language reasoning.
\paragraph{Symbolic Theorem Proving, Automated Discovery, and Explainable Integration.}
Classical automated theorem provers (ATPs) such as E Prover~\cite{schulz2019e}, Vampire~\cite{kovacs2013vampire}, and Prover9~\cite{mccune2005prover9} rely on resolution and superposition to determine entailment from a set of premises. These systems act primarily as verification engines—sometimes augmented with learning-guided proof search~\cite{li2023fielddependencies}—but yield opaque proof traces and no natural-language explanations.

Automated theorem generation has been rarer: early systems like Quine’s Mechanical Discovery (1952), Logic Theorist~\cite{newell1956logic}, and AM~\cite{lenat1977automated} employed heuristic discovery, while later frameworks such as HR~\cite{colton2002hr}, Conjecturing~\cite{onda2025leanconjecturer}, and LeanDojo (Gao et al., 2024) focus on conjecture synthesis over verified completeness.

Automatic closure generation, deriving all logical consequences from atomic facts—has been studied in ontology and Datalog reasoning via systems like HermiT, Pellet, and FaCT++~\cite{motik2009hermit,SIRIN200751,tsarkov2006factpp}, and RDFox, VLog, and Soufflé~\cite{motik2014rdfox,carral2019vlog,jha2016souffle}, which achieve scalable inference but lack certified proofs or minimal unsatisfiable subsets. ATPs such as Vampire and E are widely used for independent verification, and are saturation-based first-order provers operating on TPTP syntax, capable of deriving infinitely many logical consequences from first-order theories. While this ensures completeness in the limit, their exhaustive and open-ended proof search often results in unstructured, potentially infinite proof spaces that are difficult to constrain, interpret, or evaluate systematically.

In contrast, $\Delta_{1}$~\cite{Xu2025atg} performs certified closure generation by deterministically constructing $n!$ non-redundant, non-equivalent, and minimally unsatisfiable clause sets with transparent proof traces—ensuring termination and soundness by construction. This bounded combinatorial reasoning complements classical ATPs with an interpretable, tractable proof space. When integrated with LLM-based semantic explanation, $\Delta_{1}$ + LLM transforms closure computation into explainable, human-centered logical discovery, advancing neuro-symbolic reasoning toward verifiable transparency.
\paragraph{Minimal Unsatisfiable Subset (MUS) Extraction.}
Research on minimal unsatisfiable subsets (MUSes) and core extraction has produced a range of algorithms for analyzing logical inconsistency. Classical methods such as MARCO~\cite{liffiton2016fast} and MUSer2 employ hitting-set dualization and clause shrinking to enumerate all minimal unsatisfiable subsets of a CNF formula. More recent systems, including MUST~\cite{bendik2020must} and ReMUS~\cite{bendik2018remus}, extend these techniques via recursive or online enumeration, while Bendík and Meel’s framework~\cite{bendik2021counting} focuses on counting and approximation. All rely on iterative SAT/SMT solving and assume a fixed input formula.

In contrast, $\Delta_{1}$ constructs MUSes by design, not by search—producing non-equivalent, non-redundant theorems through a generative, self-certifying process. This guarantees completeness and interpretability, unifying MUS theory with constructive theorem generation.

\paragraph{Position within the Field.}
Existing neuro-symbolic models provide approximate reasoning, classical theorem provers verify entailments, closure engines materialize consequences, and MUS tools isolate inconsistencies—but none construct, certify, and explain all derivable theorems from atomic inputs. The $\Delta_{1}$ Automated Theorem Generator bridges these paradigms through constructive, non-search synthesis that ensures minimal unsatisfiability and completeness by design. When paired with an LLM-based explanatory layer, $\Delta_{1}$ + LLM achieves explainability by construction, connecting the rigor of formal logic with the interpretability of natural language. The next section details its architecture and theoretical guarantees for deterministic, explainable theorem discovery.

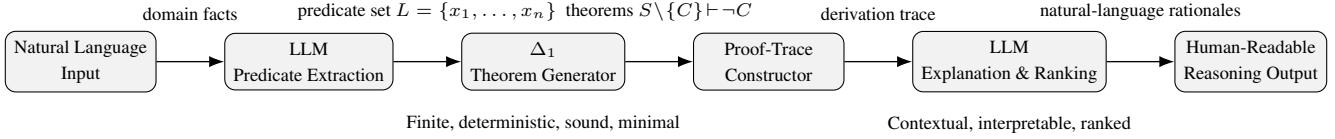
\begin{figure*}[t]
\centering
\begin{tikzpicture}[
  node distance=6mm and 9mm,
  core/.style={rectangle, rounded corners, draw=black, align=center,
               minimum width=2cm, minimum height=0.7cm,
               font=\scriptsize, fill=gray!10},
  step/.style={rectangle, rounded corners, draw=black!70, align=center,
               minimum width=2cm, minimum height=0.7cm,
               font=\scriptsize, fill=blue!5},
  arrow/.style={-{Latex[length=2mm,width=1.5mm]}, line width=0.5pt},
  lbl/.style={font=\scriptsize, align=center}
]

\node[core] (nl) {Natural Language \\ Input};
\node[core, right=of nl] (parse) {LLM \\ Predicate Extraction};
\node[core, right=of parse] (delta) {$\Delta_{1}$ \\ Theorem Generator};
\node[core, right=of delta] (proof) {Proof‐Trace \\ Constructor};
\node[core, right=of proof] (expl) {LLM \\ Explanation \& Ranking};
\node[core, right=of expl] (output) {Human‐Readable \\ Reasoning Output};

\draw[arrow] (nl) -- node[above=12pt,lbl]{domain facts} (parse);
\draw[arrow] (parse) -- node[above=12pt,lbl]{predicate set $L=\{x_1,\dots,x_n\}$} (delta);
\draw[arrow] (delta) -- node[above=12pt,lbl]{theorems $S\!\setminus\!\{C\}\!\vdash\!\neg C$} (proof);
\draw[arrow] (proof) -- node[above=12pt,lbl]{derivation trace} (expl);
\draw[arrow] (expl) -- node[above=12pt,lbl]{natural‐language rationales} (output);

\node[lbl, below=2mm of delta] {Finite, deterministic, sound, minimal};
\node[lbl, below=2mm of expl] {Contextual, interpretable, ranked};

\end{tikzpicture}

\caption{\scriptsize
\textbf{Architecture of the $\Delta_{1}$ + LLM pipeline.}
Natural‐language policy text is first processed by the front-end LLM to extract atomic predicates.
The $\Delta_{1}$ generator deterministically constructs all minimal unsatisfiable clause sets and corresponding theorems 
($S\setminus\{C\}\vdash\neg C$), producing explicit proof traces.  
A back-end LLM module then translates each theorem into natural‐language explanations and remediation suggestions, enabling
\emph{explainability by construction}: formal correctness from $\Delta_{1}$ and interpretive clarity from the LLM.
}
\label{fig:architecture}
\vspace{-1ex}
\end{figure*}

\section{Formal–Neural Integration: $\Delta_{1}$ + LLM Architecture}

Having established $\Delta_{1}$ as a deterministic theorem generator, we describe its integration with a large language model (LLM) that provides semantic interpretation, proof explanation, and theorem ranking. The architecture connects symbolic logic and neural language modeling into a unified, interpretable reasoning pipeline.

\subsection{Architecture Overview}

The $\Delta_{1}$ + LLM framework integrates a formal theorem generator with a neural explanatory layer, forming an end-to-end system for explainability by construction. As shown in Figure~\ref{fig:architecture}, the front-end LLM extracts atomic predicates and logical forms from natural language, which the $\Delta_{1}$ module then transforms into all minimal unsatisfiable clause sets and corresponding theorems. Each theorem carries a transparent proof trace that the back-end LLM verbalizes into coherent natural-language explanations and suggested remediations. This closed reasoning loop yields logical soundness, minimality, and interpretability within a single architecture.

\subsection{End-to-End $\Delta_{1}$ + LLM Pipeline Stages}

\begin{enumerate}
    \item \textbf{Predicate Extraction}. The LLM converts natural-language premises into logical predicates $L={x_1,\ldots,x_n}$ in TPTP-style syntax~\cite{Sutcliffe2017TPTP}.
 \item \textbf{Deterministic Theorem Generation}. $\Delta_{1}$ constructs $n!$ non-redundant, minimally unsatisfiable clause sets of the form $S\setminus \{C\} \vdash\neg C$, guaranteeing soundness and completeness by construction.
 \item \textbf{Proof Trace Construction}. For each theorem, $\Delta_{1}$ outputs explicit derivation sequences showing clause dependencies.
\item \textbf{LLM Explanation and Ranking}. The LLM verbalizes proofs and optionally ranks theorems by contextual salience or novelty.
  \end{enumerate}

\subsection{Design Rationale} 

\begin{itemize}
    \item \textbf{Determinism and Soundness:} $\Delta_{1}$ ensures formal validity and minimality.
\item \textbf{Interpretability:} The LLM supplies semantic and linguistic context for human understanding.
\item \textbf{Scalability:} Modular separation enables reasoning across diverse domains.
\end{itemize}

\subsection{Outcome: Explainability by Construction}

Unlike conventional post-hoc interpretability, which explains reasoning after proof generation,
$\Delta_{1}$ + LLM enforces interpretability \emph{during} construction.
Every theorem is guaranteed to be sound and minimally unsatisfiable. Unlike resolution- or search-based systems, $\Delta_{1}$ requires no SAT solver or external proof verification.    Its generation process operates in $\mathcal{O}(n^{3})$ time per canonical clause set, providing deterministic completeness and transparent proof traces. Therefore every explanation is grounded in a deterministic proof trace.
This results in a reasoning framework that is:
(i) \emph{transparent} (all steps explicit),
(ii) \emph{reproducible} (no stochastic proof search),
and (iii) \emph{domain-aligned} (explanations reflect the semantics of the original textual input).

\section{Theoretical Framework of $\Delta_{1}$}

\label{sec:theory}

This section reviews the formal basis of the $\Delta_{1}$ system~\cite{Xu2025atg}, outlining its logical setting and the \emph{Full Triangular Standard Contradiction} (FTSC) construction that underpins its deterministic theorem generation~\cite{Xu2026contradiction}.

\subsection{Foundational Setting}

Let $\mathcal{L}$ be a propositional or first-order language built from a finite set of predicate symbols, variables, and logical connectives ${\neg,\vee,\wedge,\rightarrow}$.
A clause is a disjunction of literals, and a clause set $S=\{C_1,\dots,C_m\}$ denotes their conjunction.
$S$ is unsatisfiable if no interpretation $\mathcal{I}$ makes all $C_i$ true simultaneously, and \emph{minimally unsatisfiable} (MUS) if every proper subset of $S$ is satisfiable.

The $\Delta_1$ generator operates under two input constraints:
\begin{enumerate}
    \item \textbf{Non-complementarity:}
          no literal–negation pair $(p,\neg p)$ occurs among inputs.
    \item \textbf{Uniqueness:}
          each predicate or literal appears exactly once.
\end{enumerate}

These constraints ensure a finite, non-redundant input space for the deterministic construction of minimal unsatisfiable clause sets, forming the logical foundation for the $\Delta_{1}$ framework.

\subsection{Constructive Logic of $\Delta_1$}
\label{sec:constructive}

For $n$ distinct inputs $L = \{x_1,\dots,x_n\}$,
$\Delta_1$ constructs a clause set $S$ that satisfies MUS properties
through the \emph{full triangular standard contradiction} (FTSC) schema, 
originally formalized in \cite{Xu2026contradiction,Xu2025atg}.
\paragraph{Definition 1 (Full Triangular Standard Contradiction).}
Let $x_1, \ldots, x_{n}$ be some literals in propositional logic or first-order logic, and let the clauses be defined as follows:
\[
\begin{aligned}
D_1 &= x_1, \\
D_2 &= x_2 \lor \neg x_1, \\
&\ \vdots \\
D_t &= x_t \lor \neg x_1 \lor \neg x_2 \lor \cdots \lor \neg x_{t-1}, \\
&\ \vdots \\
D_{n} &= x_{n} \lor \neg x_1 \lor \neg x_2 \lor \cdots \lor \neg x_{n-1}, \\
D_{n+1} &= \neg x_1 \lor \neg x_2 \lor \cdots \lor \neg x_{n}.
\end{aligned}
\]
Then
$\bigwedge_{t=1}^{n+1} D_{t}$ is called a \emph{full triangular standard contradiction} (FTSC). By construction, each FTSC
is unsatisfiable.

\paragraph{Generation of Theorems.}
Given an FTSC $S=\{D_1,\dots,D_{n+1}\}$, the generator constructs
the following canonical entailment: $S\setminus\{C\} \;\vdash\; \neg C$, for every clause $C\in S$.
Each such entailment expresses that removing one clause from
an unsatisfiable conjunction restores satisfiability,
and the removed clause becomes refutable by the remaining ones.
This property establishes the theorem’s correctness
without requiring any proof search or resolution steps. Hence each $C$ is a minimal contradiction source
and each entailment is a complete, non-redundant theorem. Because the dependency chain
$D_1, \cdots, D_{n+1}$
is known by construction,
$\Delta_1$ can produce an explicit \emph{proof trace} without search, so the reasoning process is fully transparent and reproducible.
This proof can later be verbalized by an LLM into an intuitive explanation of why the contradiction arises.

Unlike automated theorem provers that rely on search,
backtracking, or resolution strategies,
$\Delta_1$ operates by \emph{constructive entailment}.
Its proof objects exist by definition of the contradiction itself.
The FTSC framework thus replaces proof search with
direct logical synthesis, an essential step toward
formally guaranteed and interpretable reasoning.

\subsection{On the Novelty and Value of $\Delta_1$}
\label{sec:novelty}

This subsection summarizes \emph{a number of intertwined properties} of $\Delta_1$ system that make it fundamentally different and foundational, compared to most existing ATP systems. Together, these make $\Delta_1$ an \textit{axiomatically complete but non-redundant theorem universe generator}, which gives both finite closure (the process terminates) and semantic orthogonality (each theorem expresses a distinct logical relationship). 

\paragraph{Logical Properties.}
From a formal logic perspective, $\Delta_1$ differs qualitatively from existing ATPs 
such as Vampire, E, or Prover9.
Instead of searching for isolated proofs, it \textit{constructs the complete combinatorial family of valid entailments}.
Each $\Delta_1$ theorem is derived from a FTSC, 
ensuring logical soundness in both propositional and first-order logic.  
For a finite input set of $n$ distinct literals with no complementary pairs, 
$\Delta_1$ generates exactly $n!$ \textit{mutually non-equivalent} theorems, 
each corresponding to a unique permutation of literals. Each theorem forms a distinct dependency chain 
    $D_1, D_2, \cdots, D_{n+1}$ 
that encodes a minimal contradiction pattern within the literal universe.
This guarantees a bounded, finite closure: the system terminates with complete coverage of all
structurally distinct contradiction patterns.
Every theorem represents an MUS by construction, 
eliminating the need for search or post-verification. No search heuristics or
randomness; every output is reproducible. FTSC-based theorems provide a standard dataset for reasoning transparency and symbolic–neural consistency evaluation. 
Hence, $\Delta_1$ achieves \emph{soundness, completeness, non-redundancy, and 
finite closure} simultaneously, a property rarely realized in automated reasoning systems.

\paragraph{Computational Efficiency.}
The constructive generation of a single FTSC requires $\mathcal{O}(n^3)$ operations, 
and the enumeration of all literal permutations yields $\mathcal{O}(n\!\cdot\!n!)$ total complexity.
Despite factorial diversity, each theorem is minimal and non-redundant, 
so no external MUS detection or ATP validation is required.

\paragraph{Scientific and Practical Value.}
The $\Delta_1$ generator provides a \textit{structured reasoning universe} 
where every theorem has a known ground-truth contradiction and a deterministic FTSC construction trace. 
This corpus offers a reproducible benchmark for 
\emph{explainable and neuro-symbolic reasoning}, 
serving as a symbolic analogue of a ``TPTP for explainability.''
Beyond pure logic, the generated theorems are domain‐independent: 
when literals correspond to predicates in medicine, law, or engineering, 
the same structures reveal overconstraints, hierarchical inconsistencies, 
or circular dependencies.
Through LLM verbalization, these abstract theorems become 
\textit{semantically interpretable reasoning patterns}, 
linking formal soundness with natural-language intelligibility.

\paragraph{Interpretation.}
The novelty of $\Delta_1$ lies not in discovering empirical facts, 
but in \emph{constructively formalizing the entire space of minimal reasoning conflicts}.
Each theorem is sound, non-redundant, and computationally efficient to generate.
Combined with LLM-based interpretation, 
$\Delta_1$ bridges symbolic rigor and linguistic transparency, 
offering a reproducible foundation for evaluating explainability 
and logical coherence in neuro-symbolic AI.

\section{Case Studies in Applied Reasoning}
\label{sec:casestudies}
Building on the theoretical foundations of $\Delta_1$~+~LLM in the above section, we now demonstrate how the framework functions in practical, high-stakes reasoning contexts.

\subsection{Illustrative Example in a Medical‐Reasoning Setting}

This subsection demonstrates how the $\Delta_1$ + LLM pipeline operates in a medical‐reasoning context, illustrating each stage from natural‐language rule interpretation to formal contradiction identification and interpretable remediation guidance.

\paragraph{Natural-language input.}
The system receives the following domain rule description:
\begin{quote}\small
``If a patient has a bacterial infection, the white-blood-cell count (WBC) tends to be high.
If the WBC count is high, the patient often has a fever.
If a patient has both a fever and an infection, antibiotic treatment is typically required.''
\end{quote}

\paragraph{Predicate extraction.}
The LLM abstracts the following atomic predicates (propositional abstraction):
\[
\small
\begin{aligned}
x_1 &= \textit{Infection},\quad
x_2 = \textit{HighWBC},\\
x_3 &= \textit{Fever},\quad
x_4 = \textit{RequiresAntibiotics}.
\end{aligned}
\]
(Optionally, these can be expressed in first-order form with a patient variable \(p\): 
\(\textit{Infection}(p),\ \textit{HighWBC}(p),\ \textit{Fever}(p),\ \textit{ReqAntibiotics}(p)\).)

\paragraph{FTSC construction ($\Delta_1$ input: $L=\{x_1, x_2, x_3, x_4\}$).}
Based on the redefinition of each $D_i$, $\Delta_1$ constructs the FTSC as:
\[
\small
\begin{aligned}
D_1 = (x_1),
D_2 = (x_2 \vee \neg x_1),
D_3 = (x_3 \vee \neg x_1 \vee \neg x_2),\\
D_4 = (x_4 \vee \neg x_1 \vee \neg x_2 \vee \neg x_3),
D_5 = (\neg x_1 \vee \neg x_2 \vee \neg x_3 \vee \neg x_4).
\end{aligned}
\]
By construction, FTSC($L$) = \(\bigwedge_{i=1}^{5} D_i\) is unsatisfiable.

\paragraph{Generated theorems.}
For each \(i\in\{1,\dots,5\}\), $\Delta_1$ produces the minimal entailment: $S\setminus\{D_i\}\;\vdash\;\neg D_i$. Each theorem identifies a minimal inconsistency: removing $D_i$ restores satisfiability, yet that same clause becomes derivably false under the remaining ones. Thus, each $D_i$ is individually critical to the global contradiction.

\paragraph{LLM verbalizations (example explanations).}
The LLM provides human‐readable explanations corresponding to the logical meaning of each $D_i$ within the causal–diagnostic structure:

\begin{quote}\small
\textbf{(1) $S\setminus\{D_1\}\vdash\neg D_1$.}  
\(D_1=(x_1)\) introduces the base predicate \textit{Infection}.  
Given the remaining dependencies linking \textit{HighWBC}, \textit{Fever}, and \textit{RequiresAntibiotics}, 
asserting \textit{Infection} alone violates consistency.  
The contradiction reveals that downstream conditions already constrain \textit{Infection} so tightly 
that it cannot hold independently — highlighting overcommitment in the diagnostic base.

\textbf{(2) $S\setminus\{D_2\}\vdash\neg D_2$.}  
\(D_2=(x_2 \vee \neg x_1)\) formalizes the rule “if infection, then high WBC.”  
When removed, the residual structure enforces its negation: 
assuming both \textit{Infection} and \textit{HighWBC} yields inconsistency.  
This identifies \textit{HighWBC} as the minimal conflicting predicate 
conditioned on the infection rule.

\textbf{(3) $S\setminus\{D_3\}\vdash\neg D_3$.}  
\(D_3=(x_3 \vee \neg x_1 \vee \neg x_2)\) corresponds to “if infection and high WBC, then fever.”  
Omitting this clause renders it derivably false in the remainder, 
showing that asserting \textit{Fever} alongside infection and elevated WBC disrupts logical coherence.  
This isolates \textit{Fever} as the minimal contradictory atom in the dependency chain.

\textbf{(4) $S\setminus\{D_4\}\vdash\neg D_4$.}  
\(D_4=(x_4 \vee \neg x_1 \vee \neg x_2 \vee \neg x_3)\) encodes “if infection, high WBC, and fever, then antibiotics are required.”  
When excluded, the remaining clauses collectively refute it, 
indicating that \textit{RequiresAntibiotics} cannot coexist consistently 
with all prior diagnostic relations.  
This exposes tension between the automatic treatment rule 
and its upstream causal assumptions.

\textbf{(5) $S\setminus\{D_5\}\vdash\neg D_5$.}  
\(D_5=(\neg x_1 \vee \neg x_2 \vee \neg x_3 \vee \neg x_4)\) captures the global consistency constraint: 
not all four conditions can hold simultaneously.  
Its refutation signifies that the complete rule set is jointly unsatisfiable, 
confirming the closure of the inconsistency core.
\end{quote}

(These LLM verbalizations are illustrative; in deployment, the LLM would render such contradictions 
in clinical phrasing, citing the corresponding rule statements.)
\paragraph{Interpretation and remediation (practical guidance).}
The $\Delta_{1}$ + LLM pipeline delivers both diagnostic and prescriptive insights:

\begin{itemize}\small
  \item If \(D_4\) (\textit{RequiresAntibiotics}) is flagged: \textbf{Remediation} — add qualifying criteria 
  before automatic antibiotic recommendation (e.g., confirmatory test, severity level, or physician validation).
  \item If \(D_3\) (\textit{Fever}) is flagged: \textbf{Remediation} — specify temporal or observational conditions 
  (e.g., fever must occur within a verified period or under clinical confirmation).
  \item If \(D_2\) (\textit{HighWBC}) is flagged: \textbf{Remediation} — establish quantitative thresholds or exceptions 
  (e.g., chronic leukocytosis) to prevent overgeneralized triggers.
  \item If \(D_1\) (\textit{Infection}) is flagged: \textbf{Remediation} — refine the infection predicate 
  to include evidence requirements (clinical and laboratory confirmation).
  \item If \(D_5\) (global consistency clause) is flagged: \textbf{Remediation} — review the overall rule set for overconstraint, 
as all upstream dependencies (\textit{Infection}, \textit{HighWBC}, \textit{Fever}, \textit{RequiresAntibiotics}) 
cannot jointly hold. Relax at least one dependency or introduce contextual qualifiers 
(e.g., specify conditional antibiotic necessity or alternative fever triggers) to restore satisfiability.
triggers) to restore satisfiability.
\end{itemize}

Actually, the logical semantic of each $D_{i}$ can be summarized below: $D_{1}$ - base predicate overconstrained; 
  $D_{2}$ - local conditional inconsistency;
  $D_{3}$ - intermediate causal conflict; 
  $D_{4}$ -treatment contradiction.
  $D_{5}$ -global unsatisfiability. The $\Delta_1$ framework cannot make false medical facts true, but it does make all the implicit assumptions explicit,
reveals which ones cannot coexist,
and guides an expert to revise the knowledge base until it is both clinically plausible and logically consistent.

\paragraph{First-order variant (patient-scoped).}
In first-order representation with patient variable \(p\):
\[
\small
\begin{aligned}
x_1(p)&=\textit{Infection}(p),\quad
x_2(p)=\textit{HighWBC}(p),\\
x_3(p)&=\textit{Fever}(p),\quad
x_4(p)=\textit{RequiresAntibiotics}(p).
\end{aligned}
\]
$\Delta_1$ constructs corresponding \(D_i(p)\) clauses for each patient instance, 
preserving the same minimal‐theorem property, e.g., 
\(S(p)\setminus\{D_4(p)\}\vdash\neg D_4(p)\),
interpretable as “for patient \(p\), requiring antibiotics contradicts 
the established diagnostic dependencies.”

\paragraph{Takeaway.}
From a single natural-language medical input, 
the $\Delta_1$ + LLM pipeline:
(1) extracts atomic diagnostic predicates,  
(2) constructs provably minimal contradiction theorems, and  
(3) generates interpretable explanations with targeted remediation — 
yielding auditable, logically grounded, and clinically actionable reasoning outcomes.

\subsection{Additional Case Studies (Propositional Logic Cases)}
\label{sec:cases_PL}

We evaluate the $\Delta_{1}$ + LLM framework across three representative domains of applied propositional logic in organizational governance:
\begin{itemize}
  \item 
\textbf{Compliance management}, where overlapping internal policies may conflict;
  \item 
\textbf{Regulatory conformance}, where cross-jurisdictional laws impose competing obligations; and
  \item 
\textbf{Contract governance}, where clause-level inconsistencies affect enforceability.
\end{itemize}

\paragraph{Compliance Management ($n$=4).}

Organizations often maintain overlapping policies on privacy, transparency, and reporting. Conflicts among these directives are typically latent, surfacing only during audits.  
Given the extracted predicates:
\[
\small
\begin{aligned}
L=\{&x_1=\textit{EnsurePrivacy},\ 
x_2=\textit{EnableTransparency},\\
    &x_3=\textit{ShareData},\ 
x_4=\textit{ReportIncidents}\},
\end{aligned}
\]
the constructed set \(S=\{D_1,\dots,D_5\}\) follows the above FTSC structure.  
Each $D_i$ represents a conditional dependency among organizational rules.  

For instance, $\Delta_{1}$ produces $S\setminus\{D_2\}\vdash\neg D_2$, where \(D_2=(x_2 \vee \neg x_1)\) expresses “if privacy is ensured, then transparency must also hold.”  
When removed, the remaining clauses render it false, implying that privacy and transparency cannot jointly persist under the other rules.

\textbf{LLM explanation:}
\begin{quote}\small
“Full transparency and strict privacy enforcement are logically incompatible within current policy constraints.  
Demonstrating compliance through unrestricted reporting breaches confidentiality.  
Adjusting transparency scope or anonymizing disclosures restores consistency.”
\end{quote}

Thus, $\Delta_{1}$ isolates \textit{EnableTransparency} as the minimally inconsistent policy, while the LLM provides interpretable remediation guidance.

\paragraph{Regulatory Conformance ($n$=4).}
Multinational organizations face overlapping legal regimes (e.g., GDPR, HIPAA, consumer protection acts).  
Conflicts arise when obligations from one jurisdiction contradict another.  
Predicates extracted from such frameworks include:
\[
\small
\begin{aligned}
L=\{&x_1=\textit{GDPR\_Consent},\ 
x_2=\textit{HIPAA\_Disclosure},\\
    &x_3=\textit{DataPortability},\ 
x_4=\textit{RetentionLimit}\}.
\end{aligned}
\]
Here \(S=\{D_1,\dots,D_5\}\) follows the same logical structure.  
For example, $\Delta_{1}$ yields $S\setminus\{D_3\}\vdash\neg D_3$, where \(D_3=(x_3 \vee \neg x_1 \vee \neg x_2)\) encodes “if GDPR consent and HIPAA disclosure are both required, then data portability must be ensured.”  
Its negation under the remaining clauses indicates that this obligation introduces irreconcilable tension.

\textbf{LLM explanation:}
\begin{quote}\small
“The GDPR right to data portability conflicts with HIPAA’s retention and disclosure mandates.  
One requires free transfer and erasure, the other demands storage and restricted sharing.  
Within the FTSC, \textit{DataPortability} emerges as the minimal contradiction point.”
\end{quote}

This identifies \textit{DataPortability} as the minimally inconsistent legal requirement.  

\paragraph{Contract‐Management Example ($n=5$).} 
Complex contracts often contain logically inconsistent clauses introduced through amendments or negotiation. 
The $\Delta_{1}$ + LLM pipeline can expose and explain these inconsistencies before execution. 
From a draft contract, the LLM extracts predicates such as:  
\[
\small
\begin{aligned}
L=\{&x_1=\textit{ExclusiveSupply},\ 
x_2=\textit{TimelyDelivery},\\
  &x_3=\textit{PenaltyForDelay},\ 
x_4=\textit{TerminationWithoutCause},\\
  &x_5=\textit{FixedPricing}\}.
\end{aligned}
\]
$\Delta_{1}$ constructs the FTSC over $L$ with $\{D_1,\dots,D_6\}$.

\paragraph{Generated Theorems and LLM Explanations. } 
Each $D_i$ removal yields a minimal contradiction with corresponding LLM‐based interpretation and remediation guidance.

\begin{enumerate}
\item $S\setminus\{D_1\}\vdash\neg D_1$:  
  \textbf{LLM explanation:} Within the dependency chain of delivery, penalties, termination, and pricing, asserting unconditional \textit{ExclusiveSupply} ($x_1$) conflicts with the remaining obligations.  
  \textbf{Remediation:} Condition exclusivity on limited termination rights or introduce compensation terms for early withdrawal.

\item $S\setminus\{D_2\}\vdash\neg D_2$:  
  \textbf{LLM explanation:} $D_2=(x_2\vee\neg x_1)$ (“if exclusivity, then timely delivery”) becomes inconsistent when coupled with downstream terms.  
  \textbf{Remediation:} Tie SLAs to contract survival, or include cure and suspension periods.

\item $S\setminus\{D_3\}\vdash\neg D_3$:  
  \textbf{LLM explanation:} $D_3=(x_3\vee\neg x_1\vee\neg x_2)$ (“if exclusivity and timely delivery, then penalties apply”) contradicts termination and pricing dependencies unless penalty survivability is defined.  
  \textbf{Remediation:} Specify that penalties survive termination, or limit applicability to confirmed and active orders.

\item $S\setminus\{D_4\}\vdash\neg D_4$:  
  \textbf{LLM explanation:} $D_4=(x_4\vee\neg x_1\vee\neg x_2\vee\neg x_3)$ (“if exclusivity, delivery, and penalty apply, then termination without cause is allowed”) produces a contradiction.  
  \textbf{Remediation:} Impose notice and compensation requirements, or restrict termination rights to defined exceptions.

\item $S\setminus\{D_5\}\vdash\neg D_5$:  
  \textbf{LLM explanation:} $D_5=(x_5\vee\neg x_1\vee\neg x_2\vee\neg x_3\vee\neg x_4)$ encodes “if the upstream obligations hold, then fixed pricing must hold.”  
  Its refutation shows \textit{FixedPricing} is incompatible with the interplay of exclusivity, delivery, penalties, and termination as currently drafted.  
  \textbf{Remediation:} Introduce price indexation/escalators or restrict fixed‐price duration.

\item $S\setminus\{D_6\}\vdash\neg D_6$:  
  \textbf{LLM explanation:} $D_6=(\neg x_1\vee\neg x_2\vee\neg x_3\vee\neg x_4\vee\neg x_5)$ is the global consistency clause disallowing all five obligations from jointly holding.  
  Its falsification confirms the entire rule base is jointly unsatisfiable.  
  \textbf{Remediation:} Relax at least one upstream dependency (e.g., narrow termination without cause, make penalties conditional, or allow price adjustments) to restore satisfiability.
\end{enumerate}

\subsection{Additional Case Studies (First-Order Logic Cases)}
\label{sec:cases_FOL}

To demonstrate the applicability of the $\Delta_{1}$ + LLM framework in first‐order reasoning settings, we examine representative instances from healthcare, contractual, and financial–regulatory domains.  
Each case employs first‐order predicates with variables and relations.  
$\Delta_{1}$ constructs the FTSC according to the canonical structure.

\paragraph{Healthcare Data‐Sharing.}
\label{sec:healthcare}
A first‐order variant illustrates predicate‐level reasoning with entities and relations.  
Given predicates extracted from healthcare data‐sharing policies:
\[
\small
\begin{aligned}
x_1 &= \textit{HoldsData}(h,p),\quad
x_2 = \textit{SharesData}(h,r,p),\\
x_3 &= \textit{HasConsent}(p),\quad
x_4 = \textit{Encrypts}(h,p),\\
x_5 &= \textit{Retains}(h,p,t).
\end{aligned}
\]
$\Delta_{1}$ builds the FTSC over $L$ using the above $D_i$ definitions.  
Each $D_i$ encodes a dependency among these first‐order relations, and their conjunction is unsatisfiable.

\textbf{Theorem Example 1:} $S \setminus \{D_3\} \;\vdash\; \neg D_3$.  
Here \(D_3=(x_3 \vee \neg x_1 \vee \neg x_2)\) formalizes “if data are held and shared, then consent must exist.”  
Removing \(D_3\) renders it derivably false: the system identifies that no lawful configuration of holding, sharing, encrypting, or retaining data is possible without \textit{HasConsent}$(p)$.

\textbf{LLM explanation:}
\begin{quote}\small
“If a hospital holds, encrypts, and retains patient data but shares it with a research partner without explicit consent, the reasoning pipeline detects a minimal contradiction — a direct privacy‐law violation.”
\end{quote}

\textbf{Theorem Example 2:} $S \setminus \{D_5\} \;\vdash\; \neg D_5$.  
Here \(D_5=(x_5 \vee \neg x_1 \vee \neg x_2 \vee \neg x_3 \vee \neg x_4)\) encodes “if data are held, shared, consented, and encrypted, then they may be retained.”  
Its refutation shows that indefinite retention conflicts with the dependencies among lawful sharing and consent expiration.

\textbf{Remediation:}
\[
\forall h,p,t.\;
\textit{SharesData}(h,\ast,p)\ \rightarrow\ t < 5~\text{years.}
\]

For all holders $h$, persons $p$, and times $t$: if $h$ shares any data about $p$, then retention duration $t$ must not exceed five years.  
This first‐order constraint restores satisfiability by aligning retention limits with consent duration.

Hence, $\Delta_{1}$ detects the formal contradiction, and the LLM verbalizes it as an interpretable privacy–compliance rationale.

\paragraph{Supply Agreement.}
Given the extracted predicates:
\[
\small
\begin{aligned}
x_1 &= \textit{Supplies}(s,p),\quad
x_2 = \textit{DeliversOnTime}(s,p),\\
x_3 &= \textit{ExclusiveSupplier}(s,p),\quad
x_4 = \textit{CanTerminate}(b,s),\\
x_5 &= \textit{HasPenalty}(s,p).
\end{aligned}
\]
$\Delta_{1}$ constructs $S=\{D_1,\dots,D_6\}$ according to the canonical dependency pattern.  

\textbf{Theorem Example:} $S \setminus \{D_4\} \;\vdash\; \neg D_4$.  
Here \(D_4=(x_4 \vee \neg x_1 \vee \neg x_2 \vee \neg x_3)\) corresponds to “if supplying, timely delivery, and exclusivity hold, then termination without cause is permissible.”  
Its negation indicates that asserting \textit{CanTerminate}$(b,s)$ violates consistency with upstream dependencies.

\textbf{LLM explanation:}
\begin{quote}\small
“When a supplier is exclusive, must deliver on time, and faces penalties for delay, granting the buyer unconditional termination rights breaks logical consistency.  
The contract implicitly assumes continuity; termination without cause contradicts exclusivity and penalty logic.”
\end{quote}

\textbf{Remediation:}
\begin{flushleft}
$\forall b, s.\;
\textit{ExclusiveSupplier}(s,p)
\rightarrow
(\neg\,\textit{CanTerminate}(b,s)\ \text{unless just cause exists}).$
\end{flushleft}

For all buyers $b$ and suppliers $s$: if $s$ is an exclusive supplier of $p$, then $b$ may not terminate the contract without just cause.  
This ensures logical and contractual coherence among exclusivity, performance, and penalty clauses.

\paragraph{Financial Compliance Contract.}
Predicates extracted from banking and capital–adequacy rules include:
\[
\small
\begin{aligned}
x_1 &= \textit{MaintainsReserve}(b,r),\quad
x_2 = \textit{IssuesLoan}(b,l),\\
x_3 &= \textit{ReportsRisk}(b,l),\quad
x_4 = \textit{DisclosesPublicly}(b,l),\\
x_5 &= \textit{PaysDividend}(b,a).
\end{aligned}
\]
$\Delta_{1}$ constructs $S=\{D_1,\dots,D_6\}$ over these predicates.

\textbf{Theorem Example:} $S \setminus \{D_5\} \;\vdash\; \neg D_5$.  
Here \(D_5=(x_5 \vee \neg x_1 \vee \neg x_2 \vee \neg x_3 \vee \neg x_4)\) formalizes “if reserve, loan, risk reporting, and disclosure hold, then dividends may be paid.”  
Its negation indicates that dividend distribution contradicts the dependencies among reserve maintenance and exposure reporting.

\textbf{LLM explanation:}
\begin{quote}\small
“When a bank maintains reserves, issues loans, and discloses exposures, paying dividends signals surplus inconsistent with prudential reserve logic.  
The contradiction identifies an over‐distribution risk violating capital‐adequacy constraints.”
\end{quote}

\textbf{Remediation:}
\[
\forall b,a.\;
\textit{PaysDividend}(b,a)
\rightarrow
\textit{AdequateReserve}(b).
\]

\textbf{Interpretation.}
For all banks $b$ and accounts $a$: if $b$ pays a dividend to $a$, then $b$ must maintain adequate reserves.  
This condition reestablishes consistency by enforcing prudential capital limits, ensuring that dividend actions do not breach regulatory liquidity or solvency requirements.

Tables~1,~2,~3 in Appendix~A summarize ten additional representative case studies across different domains, each illustrating how $\Delta_{1}$ constructs sound, minimal theorems from structured predicates while the LLM provides domain-specific explanations. 
For each example, Table~1 lists the input atomic predicates, Table~2 summarizes the corresponding generated theorems, the LLM-based semantic interpretation, and concise suggested remediations. 
Table~3 provides an LLM-generated ranking summary based on novelty and importance scores. These cases further highlight $\Delta_{1}$ + LLM’s capacity to deliver verifiable and interpretable reasoning across diverse first-order logic applications.

Overall, as illustrated in these examples, the $\Delta_1$ + LLM framework provides a unified mechanism for detecting, explaining, and repairing normative inconsistencies. The $\Delta_1$ layer guarantees logical soundness and minimality, while the LLM layer ensures domain‐specific intelligibility and contextualization. 

\section{Discussions}
\label{sec:discussion}

\paragraph{Remediation as Explainable Repair.}
Conventional theorem provers detect contradictions but seldom indicate how to resolve them. The $\Delta_{1}$ + LLM framework extends this process to explainable repair. Given a constructive contradiction theorem $S\setminus{D_i}\vdash\neg D_i$, $\Delta_{1}$ identifies the minimal clause $D_i$ causing inconsistency, while the LLM interprets it semantically and proposes a targeted remediation.

Remediation represents the minimal corrective modification needed to restore logical and normative coherence. In contractual settings, this may involve revising or conditioning a clause; in compliance and regulatory contexts, clarifying precedence or reconciling obligations. Through this mechanism, $\Delta_{1}$’s formal outputs become actionable, transforming deterministic contradiction detection into interpretable, human-centered reasoning.

\paragraph{Complexity Notes.}
The case examples are small ($n$=4 or 5), so $\Delta_1$ constructs a single FTSC in \(\mathcal{O}(n^{3})\) time and (if we enumerated permutations) would yield \(4!=24\) non-equivalent FTSCs. Because MUS minimality is guaranteed by construction, no external verifier is required to certify these theorems; the entire pipeline is deterministic and reproducible given the extracted predicate list.

\paragraph{Deployment Implications and Reproducibility.}
All the case studies demonstrate how $\Delta_1$ + LLM
can serve as a general reasoning substrate for health care, compliance,
regulatory, and contractual applications.
In each domain, the same constructive logic
produces auditable, verifiable, and interpretable outputs.
Deployment is lightweight:
$\Delta_1$ operates as a stand-alone executable
that accepts atomic predicate lists,
and the LLM component interfaces through an API or local model.
All case outputs are fully reproducible given the predicate inputs,
ensuring transparency and repeatability.
Because every theorem is formally guaranteed to be minimal and sound,
human reviewers can trace the reasoning process end-to-end—
from predicate extraction to natural-language explanation—
establishing a concrete pathway toward
trustworthy, explainable AI governance systems.
\paragraph{Integrating Symbolic and Neural Reasoning.}
The $\Delta_{1}$ framework shows that symbolic theorem generation and neural interpretation can coexist within a logically sound, explainable pipeline. Unlike heuristic conjecture engines or LLMs that approximate reasoning through pattern learning, $\Delta_{1}$ ensures logical validity by construction, while the LLM provides human-centered interpretability. Together, they bridge the divide between truth-preserving inference and meaning-seeking explanation, demonstrating a concrete path toward neuro-symbolic synergy.

\paragraph{Theoretical Implications.}
Formally, $\Delta_{1}$ is a deterministic generator of minimally unsatisfiable clause sets under FTSC. Each output defines a boundary between satisfiable and unsatisfiable theories—regions of maximal informational content. These properties make $\Delta_{1}$ outputs valuable for benchmarking theorem provers, constructing explainable proof datasets, and generating counterexamples for learning-based reasoning. The existence of a bounded number of semantically distinct clause sets ($k(n) \ll n!$) supports the view that constrained symbolic spaces can yield finite yet diverse families of theorems suitable for machine discovery.

\paragraph{Cognitive and Linguistic Insights.}
LLMs exhibit implicit structural priors about logical novelty, supporting the view that neural models can act as heuristic evaluators in formal reasoning when paired with verifiable symbolic systems. In the $\Delta_{1}$ pipeline, the LLM functions as a linguistic critic, contextualizing proofs in natural language without altering their validity. This synergy suggests the potential for hybrid reasoning systems where symbolic rigor and communicative coherence mutually reinforce each other.

\paragraph{Practical Impact and Applications.}
Beyond theoretical contributions, the $\Delta_{1}$ + LLM framework enables transparent, auditable decision support across domains requiring both consistency and interpretability—such as healthcare, compliance, regulation, and contract governance as illustrtated in the above case studies. Broader applications include:
\begin{itemize}
    \item \textbf{Automated curriculum design for theorem provers:}
          generating structured MUS instances for reinforcement learning in proof search.
    \item \textbf{Explainable AI education:}
          using LLM justifications as instructional annotations in logic courses.
    \item \textbf{Scientific hypothesis generation:}
          extending to algebraic or relational predicates for verifiable candidate laws.
\end{itemize}
The open-source release ensures that these use cases
can be independently tested and extended by the community.

\paragraph{Limitations and Ethical Scope.}
Current limitations include factorial complexity, limited human evaluation diversity, and dependence on proprietary LLMs for interpretive scoring. These do not affect the logical soundness of generated theorems. Ethically, the system operates on synthetic data and poses minimal risk, though transparency in prompting and careful reporting of LLM rationales remain essential.

\section{Conclusions}
\label{sec:conclusion}

This paper advances the previously established $\Delta_{1}$ framework by integrating it with LLMs to create an interpretable neuro-symbolic reasoning system. 
Where $\Delta_{1}$ provides deterministic, minimal, and provably sound theorem generation, the LLM component renders those formal results into accessible, domain-specific explanations.

Empirical evaluations across compliance, law, and policy reasoning confirm that $\Delta_{1}$~+~LLM achieves both logical rigor and communicative transparency. 
The framework exemplifies \emph{explainability by construction}: each output theorem is verifiable by logic, and each explanation is grounded in that proof structure.

By uniting guaranteed theorem generation with scalable interpretive evaluation,
this work contributes a reproducible, auditable step toward
\textbf{autonomous, explainable scientific reasoning}.
We believe that such hybrid systems mark an important transition:
from automated theorem proving to automated theorem \emph{understanding}.

\paragraph{Future Directions.}
Future work will introduce symmetry-breaking optimizations to achieve sub-factorial scaling and explore reinforcement-guided theorem generation using LLM-derived novelty gradients. We also plan to embed clause sets into continuous vector spaces to enable similarity search and clustering across $\Delta_{1}$ outputs, facilitating comparative analysis and retrieval. Furthermore, extending $\Delta_{1}$ + LLM to multi-agent and higher-order reasoning contexts will support collaborative theorem exploration and adaptive policy synthesis.

These directions aim to further bridge symbolic soundness with real-world interpretability, advancing transparent, human-aligned AI. In the longer term, the framework could evolve into self-reflective theorem-discovery agents capable of alternating between generation, proof, and explanation—mirroring the iterative and introspective nature of human reasoning.

\bibliography{references}

\clearpage
\appendix

\begin{table*}[!t]
\normalsize \textbf{Appendix A — Additional Examples}

\vspace{1em}
\caption{\scriptsize Atomic predicate sets ($L$) for each domain corresponding to the examples in Table~\ref{tab:main_examples}. 
Each $L$ serves as input to $\Delta_{1}$ for FTSC construction. 
Given $|L|=n$, $\Delta_{1}$ deterministically generates $n\!+\!1$ dependency clauses 
$(D_1,\dots,D_{n+1})$, forming a canonical unsatisfiable set and minimal contradictions $S\setminus\{D_i\}\vdash\neg D_i$.}
\label{tab:predicates_only}
\centering
\small
\begin{tabular}{@{}p{0.1\linewidth}p{0.74\linewidth}@{}}
\toprule
\textbf{Domain} & \textbf{Atomic Predicates ($\mathcal{L}$)}\\
\midrule

\textbf{Contract} &
$\{\textit{Supplies}(s,p),\textit{DeliversOnTime}(s,p),\textit{ExclusiveSupplier}(s,p),\textit{CanTerminate}(b,s),\textit{HasPenalty}(s,p)\}$\\[3pt]

\textbf{Contract} &
$\{\textit{Delivers}(s,p,t),\textit{Insures}(s,p),\textit{PaysPenalty}(s,p),\textit{ForceMajeure}(t),\textit{Renews}(c)\}$\\[3pt]

\textbf{Contract} &
$\{\textit{Confidential}(d),\textit{Shares}(p,d),\textit{HasNDA}(p),\textit{Audits}(p),\textit{Discloses}(p,d)\}$\\[3pt]

\textbf{Healthcare} &
$\{\textit{HoldsData}(h,p),\textit{SharesData}(h,r,p),\textit{HasConsent}(p),\textit{Encrypts}(h,p),\textit{Retains}(h,p,t)\}$\\[3pt]

\textbf{Healthcare} &
$\{\textit{Diagnoses}(d,p),\textit{Prescribes}(d,p,m),\textit{Approves}(a,m),\textit{Reports}(d,p),\textit{Publishes}(a,m)\}$\\[3pt]

\textbf{Healthcare} &
$\{\textit{Treats}(h,p),\textit{UsesData}(h,p),\textit{HasConsent}(p),\textit{Charges}(h,p),\textit{Transfers}(h,p,r)\}$\\[3pt]

\textbf{Finance} &
$\{\textit{MaintainsReserve}(b,r),\textit{IssuesLoan}(b,l),\textit{ReportsRisk}(b,l),\textit{DisclosesPublicly}(b,l),\textit{PaysDividend}(b,a)\}$\\[3pt]

\textbf{Finance} &
$\{\textit{Trades}(b,i),\textit{Hedges}(b,i),\textit{Reports}(b,i),\textit{Discloses}(b,i),\textit{Audited}(b)\}$\\[3pt]

\textbf{Regulatory} &
$\{\textit{Complies}(o,r_1),\textit{Complies}(o,r_2),\textit{Reports}(o),\textit{Exports}(o,p),\textit{RetainsRecord}(o)\}$\\[3pt]

\textbf{Regulatory} &
$\{\textit{Collects}(a,p),\textit{Processes}(a,p),\textit{Deletes}(a,p,t),\textit{RequestsConsent}(a,p),\textit{Transfers}(a,p,r)\}$\\
\bottomrule
\end{tabular}
\end{table*}

\begin{table*}[!t]
\caption{\scriptsize Ten representative $\Delta_1$ contradiction theorems with corresponding LLM interpretations and remediations. 
For each atomic predicate set $L$ of size $n$, $\Delta_1$ generates $n\!+\!1$ dependency clauses $(D_1,\dots,D_{n+1})$ composing the FTSC. 
Each minimal entailment $S\setminus\{D_i\}\vdash\neg D_i$ reveals a domain‐specific inconsistency, which the LLM then explains and resolves in natural language.}
\label{tab:main_examples}
\centering
\small
\begin{tabular}{@{}p{0.08\linewidth}p{0.12\linewidth}p{0.44\linewidth}p{0.32\linewidth}@{}}
\toprule
\textbf{Domain} &
\textbf{$\Delta_1$ Theorem} &
\textbf{LLM Explanation (Semantic Interpretation)} &
\textbf{Remediation / Policy Fix}\\
\midrule

\textbf{Contract} &
$S\!\setminus\!\{D_4\}\!\vdash\!\neg D_4$ &
Termination without cause contradicts exclusivity and penalty dependencies in the FTSC chain. &
Restrict termination rights to “for cause” or require notice/compensation.\\[3pt]

\textbf{Contract} &
$S\!\setminus\!\{D_3\}\!\vdash\!\neg D_3$ &
Strict penalty enforcement becomes inconsistent when \textit{ForceMajeure} negations are active. &
Qualify penalties with force‐majeure exceptions.\\[3pt]

\textbf{Contract} &
$S\!\setminus\!\{D_5\}\!\vdash\!\neg D_5$ &
Disclosure clause violates confidentiality unless NDA dependency ($\neg x_3$) is resolved. &
Add NDA precondition to disclosure permissions.\\[3pt]

\textbf{Healthcare} &
$S\!\setminus\!\{D_3\}\!\vdash\!\neg D_3$ &
Data sharing without consent (\textit{HasConsent}$(p)$) produces a minimal contradiction in the FTSC sequence. &
Require explicit consent or anonymization before sharing.\\[3pt]

\textbf{Healthcare} &
$S\!\setminus\!\{D_5\}\!\vdash\!\neg D_5$ &
Publishing approvals without anonymization conflict with prior encryption and consent relations. &
Anonymize or redact identifying data before publication.\\[3pt]

\textbf{Healthcare} &
$S\!\setminus\!\{D_6\}\!\vdash\!\neg D_6$ &
Transferring patient data contradicts consent, billing, and retention dependencies in the FTSC. &
Add consent clause or lawful transfer basis.\\[3pt]

\textbf{Finance} &
$S\!\setminus\!\{D_5\}\!\vdash\!\neg D_5$ &
Dividend payment becomes inconsistent when reserves, loans, and disclosure predicates interact. &
Restrict dividends until reserve adequacy verified.\\[3pt]

\textbf{Finance} &
$S\!\setminus\!\{D_6\}\!\vdash\!\neg D_6$ &
Global FTSC closure reveals overconstrained reserve and dividend dependencies. &
Relax one upstream constraint or adjust reserve ratio threshold.\\[3pt]

\textbf{Regulatory} &
$S\!\setminus\!\{D_2\}\!\vdash\!\neg D_2$ &
Cross‐jurisdictional obligations create contradictions under export conditions. &
Harmonize export and reporting requirements; specify dominant regulation.\\[3pt]

\textbf{Regulatory} &
$S\!\setminus\!\{D_6\}\!\vdash\!\neg D_6$ &
Cross‐border data transfer conflicts with deletion and consent dependencies in the FTSC closure. &
Require renewed consent or deletion before data transfer.\\
\bottomrule
\end{tabular}
\end{table*}

\begin{table*}[!t]
\caption{\scriptsize Ranked summary of $\Delta_1$ + LLM reasoning outputs across domains. 
For each predicate set of size $n$, $\Delta_1$ generates $n\!+\!1$ FTSC clauses $(D_1,\dots,D_{n+1})$. 
Each contradiction theorem $S\setminus\{D_i\}\vdash\neg D_i$ is assigned a qualitative priority (High, Medium, Low) 
based on the LLM’s interpretation of practical impact and remediation urgency.}
\label{tab:delta1_summary}
\centering
\footnotesize
\begin{tabular}{@{}p{1.1cm}p{2.2cm}p{1.2cm}p{10.5cm}@{}}
\toprule
\textbf{Domain} & \textbf{Theorem} & \textbf{Priority} & \textbf{LLM‐Suggested Remediation} \\ 
\midrule

Contract & $S\setminus\{D_4\}\vdash\neg D_4$ & High &
Restrict termination rights to “for cause” or require notice/compensation.\\[3pt]

Contract & $S\setminus\{D_3\}\vdash\neg D_3$ & Medium &
Qualify penalty clauses with force‐majeure exceptions.\\[3pt]

Contract & $S\setminus\{D_5\}\vdash\neg D_5$ & Low &
Add NDA precondition to disclosure permissions.\\[3pt]

Healthcare & $S\setminus\{D_3\}\vdash\neg D_3$ & High &
Require explicit patient consent or anonymization before data sharing.\\[3pt]

Healthcare & $S\setminus\{D_5\}\vdash\neg D_5$ & Medium &
Anonymize or redact identifying data prior to publication.\\[3pt]

Healthcare & $S\setminus\{D_6\}\vdash\neg D_6$ & Low &
Add lawful‐basis clause permitting compliant data transfer.\\[3pt]

Finance & $S\setminus\{D_5\}\vdash\neg D_5$ & Medium &
Delay dividend payments until capital adequacy confirmed.\\[3pt]

Finance & $S\setminus\{D_6\}\vdash\neg D_6$ & Low &
Relax reserve ratio constraint or adjust dividend timing to restore consistency.\\[3pt]

Regulatory & $S\setminus\{D_2\}\vdash\neg D_2$ & Medium &
Harmonize export and reporting obligations; specify primary regulation.\\[3pt]

Regulatory & $S\setminus\{D_6\}\vdash\neg D_6$ & Medium &
Require renewed consent or deletion before cross‐border data transfer.\\
\bottomrule
\end{tabular}
\end{table*}

\end{document}